\documentclass[prl,twocolumn,floatfix,showpacs]{revtex4}
\usepackage{amsmath,amsfonts}
\usepackage{graphicx}
\newcommand{\dr}{{\rm d}}

\newcommand{\bea}{\begin{eqnarray}}
\newcommand{\beq}{\begin{equation}}
\newcommand{\eea}{\end{eqnarray}}
\newcommand{\eeq}{\end{equation}}

\begin{document}
\title
{Instabilities, nonhermiticity and exceptional points in the cranking model}
\date{\today}
\author{W.\ D.\ Heiss$^{1}$ and R.\ G.\ Nazmitdinov$^{2,3}$}
\affiliation{$^{1}$National Institute for Theoretical Physics, \\
Stellenbosch Institute for Advanced Study, \\ 
and Institute of Theoretical Physics,
University of Stellenbosch, 7602 Matieland, South Africa\\
$^{2}$Departament de F{\'\i}sica,
Universitat de les Illes Balears, E-07122 Palma de Mallorca, Spain\\
$^{3}$ Bogoliubov Laboratory of Theoretical Physics,
Joint Institute for Nuclear Research, 141980 Dubna, Russia}

\begin{abstract}
A cranking harmonic oscillator model, widely used for the physics of fast rotating 
nuclei and Bose-Einstein condensates, is
re-investigated in the context of ${\cal PT}$-symmetry. The instability points of the 
model are identified as
exceptional points. It is argued that - even though the Hamiltonian
appears hermitian at first glance - it actually is not hermitian within the
region of instability.
\end{abstract}

\maketitle

Quantum instabilities are attracting considerable attention in a
variety of physical situations. They can be associated with the formation of solitons and
vortices in Bose-Einstein condensates \cite{bc}, with a sudden change of the 
moment of inertia of a rotating nucleus (see, for example, Ref.\onlinecite{jr} 
and references therein) and a transition  from one- to two-dimensional nuclear 
rotation \cite{dr}.
A particular example of interest is the Hamiltonian
\begin{eqnarray}
\hat H&=&\hbar\omega_1 (a_{1}^{\dagger}a^{}_{1}+\frac{1}{2})
+\hbar\omega_2(a_{2}^{\dagger }a^{}_{2}+\frac{1}{2})+\nonumber\\
&&+i\hbar g_1 (a_{1}^{\dagger}a^{}_{2} - a_{2}^{\dagger}a^{}_{1})
-i\hbar g_2(a_{1}^{\dagger}a_{2}^{\dagger} - a_{2}a_{1})
\label{h1}
\end{eqnarray}
used in condensed matter physics to
describe in a simple way the interaction between an atom 
and a radiative field \cite{sh}. 
Note that the bi-linear form of (\ref{h1}) corresponds to a
linearised version of some more general interactions. As discussed below this 
may bring about an instability. Higher order terms may or may not remove such
instability.

Using the standard relations ($\hbar=m=1$)
\bea
a_k&=&\sqrt{\frac{\omega_k}{2}}x_k+i\sqrt{\frac{1}{2\omega_k}}p_k\\
a^{\dagger}_k&=&\sqrt{\frac{\omega_k}{2}}x_k-i\sqrt{\frac{1}{2\omega_k}}p_k
\eea
where $x_1=x,x_2=y$, and choosing special values for the strength constants
$g_1=\Omega(\omega_1+\omega_2)/2\sqrt{\omega_1\omega_2}$ and
$g_2=\Omega(\omega_2-\omega_1)/2\sqrt{\omega_1\omega_2}$,
one recognises the well-known cranking Hamiltonian (Routhian)
\beq 
H=\frac{p_x^2}{2}+\frac{\omega _x^2}{2}x^2
+\frac{p_y^2}{2}+\frac{\omega _y^2}{2}y^2-\Omega L_z
\label{ham}
\eeq
which has been applied in nuclear physics \cite{BM75,BR} and 
for rotating Bose-Einstein condensates (cf Ref.\onlinecite{fet}). 
The Hamiltonian (\ref{ham}) appears as the sum of
hermitian operators and is expected - naively at first glance - to be
itself a hermitian
operator. The same holds when (\ref{ham}) is written in second
quantised form (\ref{h1}).
In the following we explore the formal character of the instability
points of $H$ and $\hat H$. We argue that the operators are no longer
hermitian at these points, in fact, we show, that these points are
exceptional points (EP) \cite{kato,hesa}.

Non-hermitian Hamilton operators have attracted widely spread interest
during the recent years (see \cite{spec}), 
be it in the context of effective theories
\cite{sgh}, or in the context of finding a hermitian equivalent
\cite{mus} or in the context of ${\cal PT}$-symmetry \cite{cmb}
(${\cal PT}$ is the product of the parity and time reversal operator). One 
specific aspect of non-hermitian operators are the EPs, being singularities of 
spectrum and eigenfunctions. As such, they are usually of particular physical 
significance. They have been discussed in a great variety of physical applications:
in optics \cite{berden}, in mechanics \cite{shu}, as coalescing
resonances \cite{mond,korsch}, in atomic physics \cite{lat}, and in
more theoretical context in ${\cal PT}$-symmetric models \cite{zno}
or in considering their mutual influence \cite{sey}, to name just a few.
In its simplest case they give rise to level repulsion being
the more pronounced the nearer they lie to the real
axis. Depending on the particular situation they can signal a phase
transition \cite{hege}. In the present case the EP is associated with the onset 
of an instability. 
Note that the Hamiltonian (\ref{ham}) is symmetric under 
${\cal PT}$-operation irrespective of a special choice of 
parameters $(\Omega \rightarrow -\Omega$ under $\cal T)$ \cite{ZA}.

It is well known that a Bogoliubov transformation of the Hamiltonian (\ref{ham})
\beq
\begin{pmatrix}
q_+\cr q_-\cr q_-^{\dagger}\cr q_+^{\dagger}
\end{pmatrix}
={\cal {B}} 
\begin{pmatrix}
a_x \cr a_y \cr a_y^{\dagger} \cr a_y^{\dagger }
\end{pmatrix}
\label{bog}
\eeq
yields the form (cf \cite{BR})
\beq
{\cal {\hat H}}=\omega_+(q_+^{\dagger}q^{}_+ +\frac{1}{2}) + 
\omega_-(q_-^{\dagger}q^{}_-+\frac{1}{2})  \label{eigm}
\eeq
with the eigenmode energies
\beq
\omega_{\pm} ^2= \frac{1}{2}\big(\omega_x^2+\omega_y^2+2 \Omega
^2\pm \sqrt{(\omega_x^2-\omega_y^2)^2+8\Omega^2 (\omega_x^2+\omega_y^2)}
\;\big)  .       \label{nm}
\eeq
It is also known \cite{BR,ZA} that $\omega_- ^2$ becomes negative when the
rotational speed $\Omega $ lies between $\min(\omega_x,\omega_y)$ and
$\max(\omega_x,\omega_y)$. In the following we assume that 
$\omega_x>\omega _y$.
At the points where the two eigenmodes vanish, that is when
$\omega_{-,1}=+\omega_-$ and $\omega_{-,2}=-\omega_-$
coalesce, the matrix ${\cal B}$ in (\ref{bog}) becomes singular.
This happens at the critical points 
$\Omega _{c1} =\omega_y$ or $\Omega _{c2} =\omega_x$
signalling an instability.

The coalescence is reminiscent of the behaviour of an EP. 
To confirm that we are in fact encountering a genuine
EP and not a usual degeneracy, we have to analyse the
eigenfunctions of the respective Hamiltonians. Of course, this is
closely related to the singular behaviour of the Bogoliubov transformation
${\cal B}$.

To illuminate both, the underlying physics and the mathematical
structure, it is convenient to construct the matrix ${\cal U}$ 
connecting the original canonical coordinates $\vec p$ and $\vec r$
with the quasi-boson operators $q_k$ and $q_k^{\dagger}$, that is
\beq
\begin{pmatrix}
p_x \cr p_y \cr x \cr y
\end{pmatrix}
={\cal U}
\begin{pmatrix}
q_+\cr q_-\cr q_-^{\dagger}\cr q_+^{\dagger}
\end{pmatrix} .  \label{u}
\eeq

As a first step we aim at the generalised classical normal mode coordinates $\vec
P=(P_+,P_-)$ and $\vec X=(X_+,X_-)$ in which $H$ assumes the form
\beq
\tilde H=\frac{P_+^2}{2}+\frac{\omega_+^2}{2}X_+^2 +
\frac{P_-^2}{2}+\frac{\omega_-^2}{2}X_-^2  . \label{class}
\eeq
This is achieved by solving the classical equations of motion
\bea
\frac{\dr}{\dr t}p_k&=&-\frac{\partial H}{\partial r_k} \\
\frac{\dr}{\dr t}r_k&=&\frac{\partial H}{\partial p_k} 
\eea
which can be written in matrix form
\beq
\frac{\dr}{\dr t} \begin{pmatrix}
\vec p \cr  \vec r
\end{pmatrix} = {\cal M} \begin{pmatrix}
\vec p \cr  \vec r \end{pmatrix}    \label{eqm}
\eeq   
with
\beq
{\cal M}=\begin{pmatrix}0 & 0 & -1 & 0 \cr 0 & 0 & 0 & -1 
\cr +1 & 0 & 0 & 0 \cr 0 & +1 & 0 & 0   \end{pmatrix} {\cal H}
     \label{mat}
\eeq
where 
\beq
{\cal H} = \frac{1}{2}\begin{pmatrix}
1 & 0 & 0 & \Omega \cr 0 & 1 & -\Omega & 0 
\cr 0 & -\Omega &  \omega_x^2 & 0
  \cr \Omega & 0 & 0 & \omega_y^2  
\end{pmatrix} .
\eeq
Note that (\ref{ham}) can be written as
\beq
H=\big(\vec p \quad \vec r\big)\;{\cal H} \begin{pmatrix}
\vec p \cr  \vec r \end{pmatrix} . \label{mform}
\eeq

The solution of (\ref{eqm}) is obtained by exponentiation and reads
\beq
\begin{pmatrix}
\vec p(t) \cr  \vec r(t) \end{pmatrix} = 
{\cal U} \exp ({\cal D\,{\it t}}) {\cal V}
\begin{pmatrix}
\vec p(0) \cr  \vec r(0) \end{pmatrix} \label{sol}
\eeq
where ${\cal D}={\rm diag}(-i\omega_+,-i\omega_-,i\omega_-,i\omega_+)$ 
is the diagonal form of ${\cal M=UDV}$ 
containing the eigenmodes. The columns of ${\cal U}$ and ${\cal V}$ are the right
hand and left hand eigenvectors, respectively, of ${\cal M}$. Note
that the eigenmodes are obtained from the non-symmetric matrix ${\cal M}$; 
from this classical view point it is therefore no surprise that
some of the eigenvalues occurring in (\ref{eigm}) may be complex. As
the column vectors of ${\cal U}$ and ${\cal V}$ form a bi-orthogonal
system, we can choose ${\cal V}={\cal U}^{-1}$. Also, we observe from the
special form of (\ref{mat}) that
\beq
{\cal V}=\begin{pmatrix}0 & 0 & 0 & -i \cr 0 & 0 & -i & 0 
\cr 0 & +i & 0 & 0 \cr +i & 0 & 0 & 0 \end{pmatrix} {\cal U}^T
\begin{pmatrix}0 & 0 & +1 & 0 \cr 0 & 0 & 0 & +1 
\cr -1 & 0 & 0 & 0 \cr 0 & -1 & 0 & 0   \end{pmatrix}  . \label{inv}
\eeq

While the explicit form of $\vec P(t)$ and $\vec X(t)$ is of little interest
the essential point here is the classical instability occurring for
negative values of $\omega_-^2$, that is for
$\omega_y\le \Omega \le \omega_x$. In
fact, the harmonic oscillator potential has the 'wrong' sign in
(\ref{class}) for the
coordinates $P_-(t)$ and $X_-(t)$.
From (\ref{sol}) we read off the classical 'run away' solution in this
parameter range yielding the 
\break $\sim \exp (|\omega_-|t)$ behaviour
for position and momentum. The corresponding quantum mechanical
behaviour is discussed below.

Using the form (\ref{mform}) we aim at a form corresponding to
(\ref{eigm}), {\it viz.}
\beq
\hat H=\big(q_+ \;q_-\;q_-^{\dagger}\;q_+^{\dagger} \big)\;H_{QM} 
\begin{pmatrix} q_+ \cr q_-\cr q_-^{\dagger}\cr q_+^{\dagger}
\end{pmatrix}
\eeq
with
\beq
H_{QM}=\frac{1}{2}\begin{pmatrix}
0 & 0 & 0 & \omega_+ \cr 0 & 0 & \omega_- & 0 
\cr 0 &  \omega_- & 0 & 0 \cr   \omega_+ & 0 & 0 & 0
\end{pmatrix} .
\eeq
From (\ref{u}) this implies that ${\cal U}$ must be normalised such
that ${\cal H}={\cal U}^T H_{QM}\cal{U}$. The explicit form of the
matrix elements of ${\cal U}$ are given in \cite{Zel}, however the
the quoted paper focuses upon significantly smaller values of $\Omega $
than the range of instability.
The analytic form allows pertinent statements in
general, and in particular an expansion in $\Omega $
around the critical points $\Omega _{c1}=\omega _y$ and $\Omega_{c2}=\omega _x$.

\begin{figure}[th]
\includegraphics[height=0.14\textheight,clip]{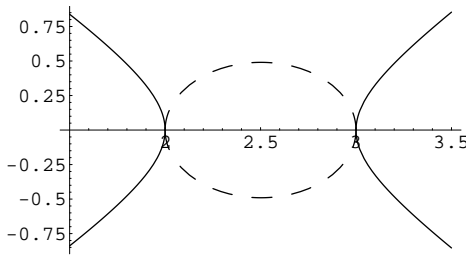}
\caption{Relevant spectrum $\pm \omega _-$ as a function of $\Omega $
  in arbitrary units.
Parameters chosen are $\omega _x=3$ and $\omega _y=2$. The dashed
lines indicate the imaginary part.}
\label{crank}
\end{figure}
The essential results are as follows: 

(i) When $\Omega \notin [\omega _y,\omega _x]$ the commutators
$[q_j^{},q_k^{\dagger}]=\delta_{j,k}$ follow from
$[r_m,p_n]=i\delta_{m,n}$. It guarantees that the boson operators
are creation and annihilation operators for the excitations, in
the present case, with
energies $\omega _-$ and $\omega _+$. However, this holds 
{\it only} when $\omega _-^2$ is positive;
if $\omega _-^2$ is negative ($\Omega \in[\omega_y,\omega_x]$) the
commutator $[q_-^{},q_-^{\dagger}]$
is negative with the implication that the operators
$q_-$ and $q_-^{\dagger}$ are no longer
proper boson operators. We further note that ${\cal V}$ - as given by
(\ref{inv}) - no longer is the inverse of ${\cal U}$ for this parameter range.

(ii) The end points of the instability region, i.e.~the points
$\Omega _{c1}=\omega_y$ and $\Omega _{c2}=\omega_x$ can be clearly
identified as EPs. In fact,
while the two eigenvectors associated with
the two distinct eigenvalues $+\omega _-$ and $-\omega _-$ are obviously
linearly independent, they become aligned, i.e.~linearly dependent, at 
$\Omega _{c1}$ and $\Omega _{c2}$ where $\omega_-$ vanishes; this is the {\it clear signature} of an
EP \cite{heha}. We recall: a genuine degeneracy would have two linearly
independent eigenvectors. EPs are a universal
phenomenon occurring in spectra and eigenfunctions under variation
of parameters. For hermitian operators they can occur only when such
parameters are continued into the complex plane thus rendering the
original hermitian operator effectively nonhermitian.

EPs are square
root singularities of the spectrum: in the present case the spectrum
has a branch cut in $\Omega $ ranging from $\omega_y$ to $\omega_x$.
When the eigenvalues $+\omega _-$ and $-\omega _-$ are continued
beyond the EP, they become imaginary (as was also noticed in \cite{ZA}) for
$\Omega \in [\omega_y,\omega_x]$, again with opposite sign (see Fig.1); 
clearly this contradicts $H$ being hermitian for this parameter range.

(iii) The correct normalisation enforced by (\ref{u}) (to guarantee
the correct commutation relations when $\Omega \notin [\omega_y,
\omega_x]$) has the consequence that
the leading terms of the components of the critical eigenvectors
behave as $(\Omega -\Omega _c)^{-1/4}$
when approaching the critical point. This particular singular
behaviour - the forth root and the infinity - is again a consequence
of the eigenfunctions at an EP \cite{hei}. In fact, it has been
shown in general \cite{dem} that the scalar product of the two eigenfunctions -
associated with the two coalescing levels - 
must vanish as a square root, in the present case
as $(\Omega -\Omega _c)^{1/2}$. As a consequence, when normalisation
is enforced by dividing by the square root of the scalar product, the singular 
behaviour follows as indicated. Moreover, the forth root
has the consequence that - for the wave function - a clockwise
encircling of the EP in the $\Omega $-plane 
yields a result that has a phase that is different from 
that of a counterclockwise encirclement. In
fact, considering $\root 4 \of z$ (taking $z=\Omega-\Omega _c$), 
one obtains $+i$ when $z$ has
described a full counterclockwise circle around zero and $-i$ when going in the
opposite direction.
This particular Riemann sheet structure has been experimentally established in microwave
cavities \cite{darm}. It would be a challenge to confirm it in the present context with
a BEC or with Raman scattering using an incident laser beam upon vibrational modes of
a medium.

So far, we have established the seemingly surprising result that the
Hamilton operator (\ref{ham}) - or its second quantised counterpart -
fails to be hermitian when $\omega _y \le \Omega \le
\omega _x$. The endpoints of this interval are EPs. We stress that
this result is based on an analytic continuation obtained from the
range $\Omega < \omega _y$, or equally, from the range $\Omega >
\omega _x$. These two (hermitian) ranges are of course also
analytically connected.

It appears apposite to contrast our findings with common wisdom about
the solutions of the Schroedinger equation of (\ref{class}). In fact,
if (\ref{class}) is considered in isolation, the Hamiltonian appears
perfectly hermitian, also for $\omega _-^2<0$. It has a continuous
spectrum associated with the unbounded classical  motion in the coordinates
$P_-$ and $X_-$. The quantum mechanical wave function is
asymptotically of the form $\exp (i |\omega _-|X_-^2/2)$ apart
from a hypergeometric function. The crucial aspect explaining this
apparent discrepancy lies in the transformation that brings us from
(\ref{ham}) to (\ref{class}). As long as $\Omega \notin [\omega_y,
\omega_x]$ the two operators are equivalent up to a similarity 
transformation. For $\Omega \in [\omega_y,
\omega_x]$ they are not. And the transformation breaks down exactly at
the EPs, the singularity that signals the instability point.

This is a beautiful demonstration of a ${\cal PT}$-symmetric operator \cite{cmb},
yet with a special twist: (\ref{ham}) appears hermitian to the naked
eye, but its spectrum is not real when $\Omega \in
[\omega_y,\omega_x]$. While the operator is  ${\cal
PT}$-symmetric, the symmetry is broken by the state vector. 
Thus, in this parameter range the hermitian form
(\ref{class}) is not its hermitian equivalent.

In conclusion, we mention that the two exceptional points collapse into
a diabolic point \cite{mvb} when $\omega_y=\omega_x$; in this case
$\Omega=\omega_x$ is a regular point with a genuine degeneracy for $\omega_-=0$.

\section*{Acknowledgements}
We are grateful for a critical reading and helpful comments by Hendrik
B Geyer.
R.G.N. is thankful for the warm hospitality which he received
from the Department of Physics of Stellenbosch University during his
visit to South Africa.
This work is partly supported by JINR-SA Agreement on scientific collaboration, 
Grant No. FIS2005-02796 (MEC, Spain) and Ram\'on y Cajal programme (Spain).

\end{document}